\documentclass[aps,prd,onecolumn,showpacs,nofootinbib,amsmath,amssymb,floatfix,superscriptaddress,showkeys]{revtex4}
\usepackage{graphicx}% Include figure files
\usepackage{dcolumn}% Align table columns on decimal point
\usepackage{bm}% bold math
\usepackage{captcont}
\usepackage{xcolor}
\usepackage{supertabular}

\usepackage{epstopdf}
\usepackage{mathtools}
\usepackage{natbib}
\usepackage{ulem}
\usepackage[colorlinks,linkcolor=red,anchorcolor=blue,citecolor=blue]{hyperref}

\newcommand{\jcap}{J. Cosmol. Astropart. Phys.}
\newcommand{\apjl}{Astrophys. J. Lett.}
\newcommand{\physrep}{Phys. Rep.}
\newcommand{\mnras}{Mon. Not. R. Astron. Soc.}
\newcommand{\araa}{Annu. Rev. Astron. Astrophys.}
\newcommand{\aap}{Astron. Astrophys}
\newcommand{\aj}{Astron. J.}
\newcommand{\plb}{Phys. Lett. B}

\begin{document}

\title{Search for gamma-ray line features from Milky Way satellites with Fermi-LAT Pass 8 data}
\author{Yun-Feng Liang}
\affiliation{Key Laboratory of Dark Matter and Space Astronomy, Purple Mountain Observatory, Chinese Academy of Sciences, Nanjing 210008, China}
\affiliation{University of Chinese Academy of Sciences, Beijing, 100012, China}
\author{Zi-Qing Xia}
\affiliation{Key Laboratory of Dark Matter and Space Astronomy, Purple Mountain Observatory, Chinese Academy of Sciences, Nanjing 210008, China}
\affiliation{School of Physics, University of Science and Technology of China, Hefei, 230026, China}
\author{Zhao-Qiang Shen}
\email{zqshen@pmo.ac.cn}
\affiliation{Key Laboratory of Dark Matter and Space Astronomy, Purple Mountain Observatory, Chinese Academy of Sciences, Nanjing 210008, China}
\affiliation{University of Chinese Academy of Sciences, Beijing, 100012, China}
\author{Xiang Li}
\email{xiangli@pmo.ac.cn}
\affiliation{Key Laboratory of Dark Matter and Space Astronomy, Purple Mountain Observatory, Chinese Academy of Sciences, Nanjing 210008, China}
\author{Wei Jiang}
\affiliation{Key Laboratory of Dark Matter and Space Astronomy, Purple Mountain Observatory, Chinese Academy of Sciences, Nanjing 210008, China}
\affiliation{School of Physics, University of Science and Technology of China, Hefei, 230026, China}
\author{Qiang Yuan}
\affiliation{Key Laboratory of Dark Matter and Space Astronomy, Purple Mountain Observatory, Chinese Academy of Sciences, Nanjing 210008, China}
\author{Yi-Zhong Fan}
\email{yzfan@pmo.ac.cn}
\affiliation{Key Laboratory of Dark Matter and Space Astronomy, Purple Mountain Observatory, Chinese Academy of Sciences, Nanjing 210008, China}
\author{Lei Feng}
\affiliation{Key Laboratory of Dark Matter and Space Astronomy, Purple Mountain Observatory, Chinese Academy of Sciences, Nanjing 210008, China}
\author{En-Wei Liang}
\affiliation{Guangxi Key Laboratory for the Relativistic Astrophysics, Department of Physics, Guangxi University, Nanning 530004, China}
\author{Jin Chang}
\affiliation{Key Laboratory of Dark Matter and Space Astronomy, Purple Mountain Observatory, Chinese Academy of Sciences, Nanjing 210008, China}

\date{\today}

\begin{abstract}
With 91 months of the publicly available Fermi-LAT Pass 8 data, we analyze the gamma-ray emission from the Milky Way satellites to search for potential line signals due to the annihilation of dark matter particles into double photons. The searched targets include a sample of dwarf spheroidal galaxies, the Large Magellanic Cloud (LMC) and Small Magellanic Cloud (SMC).
No significant line emission has been found neither in the stacked dwarf galaxy sample nor in the direction of LMC/SMC. The corresponding upper limits on the cross section of DM annihilation into two photons are derived. Compared with results of previous gamma-ray line searches with the Pass 7 data, the current constraints on the line emission from dwarf spheroidal galaxies has been significantly improved in a wide energy range. With the rapid increase of the sample of dwarf spheroidal galaxies (candidates), we expect that the sensitivity of gamma ray line searches will be significantly improved in the near future.
\end{abstract}


\maketitle

\section{Introduction}
Many astrophysical and cosmological phenomena, such as the discrepancy between luminosity masses and kinematic masses of galaxy clusters, the flat rotation curves of galaxies, and the cosmic microwave background power spectrum, indicate the existence of a large amount of so-called dark matter (DM) in the Universe. Though it is well established that the DM consists of $\sim 26\%$ of the total energy density of the current Universe, its nature is still far from clear since all the evidence/properties are inferred from gravitational effects. It is highly necessary to find non-gravitational evidence of the DM.
One way is to identify the annihilation or decay products of DM (i.e., the DM indirect detection), including photons, electrons/positrons, protons/anti-protons, and neutrinos/anti-neutrinos. Weakly Interacting Massive Particles (WIMPs) are the most extensively-investigated candidates. If they annihilate or decay, GeV-TeV gamma-rays and cosmic rays are generated. These signals, with distinct spectrum and/or spatial distribution, are among the key targets of many ground or space based instruments, including for instance Fermi-LAT \cite{atwood09lat,charles16review}, AMS-02 \cite{ams13positron}, HESS \cite{hess14dsph} and IceCube \cite{icecube15dm}. Given positive detections, the DM mass $m_\chi$ and the velocity averaged annihilation cross section ${\langle{\sigma}v\rangle}$ or the lifetime of DM particles can be reliably inferred.

%If DM particles are WIMPs, since they can engage in the weak interaction, they could annihilate or decay into standard model particles such as gamma-rays, electrons, positrons and so on which can be detected by many ground and space based instruments, i.e., Fermi-LAT, AMS2, VERITAS, IceCube. Through detecting these productions, one can derive the properties, such as velocity averaged cross section ${\langle{\sigma}v\rangle}$ and mass, of DM.

Since the launch of Fermi satellite in 2008 \cite{atwood09lat}, dedicated efforts have been made to search for the continual gamma-ray radiation from the DM annihilation or decay towards various targets, for example the dwarf spheroidal galaxies (dSphs) \cite[e.g.,][]{fermi11dsph,Geringer-Sameth11dsph,tsai13dsph,fermi14dsph,fermi15dsph,hooper15ret2,gs15ret2, des_fermi15dsph,gs15dsph,li16dsph,fermi2016dsph}, galaxy clusters \cite[e.g.,][]{yuan10gcls,huang12cluster,ando12fornax,fermi15virgo}, the Galactic center \cite[e.g.,][]{hooper2011,gordon13gc,hooper13gc,calore15gc,zb15gc}, the Large/Small Magellanic Cloud (LMC/SMC) \cite{caputo16smc,buckley15lmc} and Smith high-velocity cloud \cite{dw14smith}.
So far no credible DM signal has been identified \citep[see][for a recent review]{charles16review} and the most stringent constraint on the cross section of DM annihilating into quarks and leptons comes from the joint analysis of fifteen dSphs, which rules out the standard thermal relic cross section up to 100GeV \cite{fermi15dsph}.
This benefits from their proximity, large DM content, and low astrophysical background of the targets, as well as the sensitivity improvement by combining multiple dSphs together using a joint likelihood technique. All these merits make dSphs the most ideal objects to search for DM.

Besides the continual gamma-ray signal, DM particles can directly annihilate into monochromatic gamma-rays, i.e., via $\chi\chi \rightarrow {\gamma}X$, where $X$ could be $\gamma$, $Z$ or $h$.
The energy of the monochromatic gamma-ray line is expected to be $E_\gamma=m_\chi(1-m_X^2/4m_\chi^2)$. Though the line signal from DM annihilation would be generally very weak due to loop suppression, it is a smoking gun compared to the continual signal since the regular astrophysical processes can not generate such a signal \cite{charles16review}. The continual signal, on the other hand, suffers from significant contamination of the astrophysical background. Shortly after Fermi's launch, the Fermi-LAT collaboration had searched for line signal from DM annihilation or decay using the first 11 months' data in the energy range of 20-300 GeV \cite{fermi10line1}.
Subsequently, Fermi-LAT data have been continually re-analyzed by Fermi collaboration with more and more data accumulated, and improved analysis approach towards different Regions of Interests (ROIs) \cite{fermi12line2,fermi13line3,fermi14line4,fermi15line5}.
None of these searches found significant line signal and constraints on the annihilation cross section or lifetime of DM particles have been derived. In 2012, Bringmann et al.\cite{bringmann130gev} found a tentative line-like excess around 130 GeV when searching for internal bremsstrahlung signal from the Galactic center. Later on, this 130 GeV line was supported by several independent analyses of the gamma-ray emission in the directions of the Galactic center and some galaxy clusters \cite{weniger130gev,Sumeng2012,tempel130gev}.
However, the finding of such a line-like emission in the Earth's limb emission makes it looks more like a systematic error \cite{bloom13earthlime}. In the newly released Pass 8 data such a signal becomes negligible \cite{fermi15line5}. A recent analysis on galaxy clusters also yields null result on this 130~GeV line signal \cite{anderson16gcline}. Recently, Liang et al.\cite{liang16line} (hereafter L16) searched for gamma-ray line signals in the directions of 16 nearby galaxy clusters and found an unexpected line-like structure at $\sim$43~GeV. The  global significance of such a signal, however, is relatively low (see \cite{feng2016} for possible interpretations).

As mentioned above, dSphs are the ideal targets to search for DM signals. Indeed, the continual gamma-ray emission of DM annihilation in dSphs has been extensively searched \cite{fermi11dsph,Geringer-Sameth11dsph,tsai13dsph,fermi14dsph,fermi15dsph,hooper15ret2,gs15ret2, des_fermi15dsph,gs15dsph,li16dsph,fermi2016dsph}.
However,  the gamma-ray line signals had just been examined with the first 4-year Fermi-LAT Pass 7 data in the directions of seven targets  \cite{huang12130gev,gs12dsphline}.
In the past few years, the number of confirmed dSphs and candidates has increased rapidly. On the other hand, Fermi-LAT has collected much more data with a significantly improved reconstruction quality. In particular the latest Pass 8 data with subgroups based on the reconstruction energy resolution are very suitable for gamma-ray line searches. Therefore it is the time to re-search for line-like signal from dSphs. In this work, we adopt Fermi-LAT 91 month Pass 8 data to carry out the analysis (both the samples of confirmed dSphs and dSph candidates would be discussed).
The LMC and SMC, two other satellite galaxies of the Milky Way, are also investigated separately (see Sec. IV).
%For completeness, we will also explore whether there is any monoenergetic line signal from these two sources (see Sec. IV).

\section{TARGET DSPHS AND $J$-FACTORS}
\label{samples}
The dSphs of the Milky Way are characterized by large mass to luminosity ratios, indicating the presence of large amount of non-baryonic matter, and thus are ideal targets to search for the DM annihilation signals. However, due to their low luminosity, only $\sim$10 classical dSphs were discovered before the Sloan Digital Sky Survey (SDSS) \cite{mateo98dsphrv}.
Benefited from the long-term wide-field searches of the SDSS, this number increased to 25 over the last decade \cite{mcconnachie12}.
In the past two years, new wide-field optical imaging surveys, such as the Dark Energy Survey (DES) and the Panoramic Survey Telescope and Rapid Response System 1 (Pan-STARRS1) 3$\pi$ survey
discovered more than 20 new objects with photometric characteristics similar to the known dSphs \cite{des15y1,des15y2,laevens15tri2,laevens15_3,kim15pegasus3}.

The expected gamma-ray flux from the DM annihilation can be expressed as \cite{charles16review}
\begin{equation}
\frac{d\phi}{dE} = \phi^{\rm PP}(E_{\gamma}){\times}J_{\rm anni},
\label{equ_dmflux}
\end{equation}
where the first term is related to the particle physics and takes the form of
\begin{equation}
\phi^{PP}(E_{\gamma})=\frac{1}{2}\frac{\langle{\sigma}v\rangle}{4{\pi}m_{\chi}^2}\frac{dN_{\gamma}}{dE_{\gamma}},
\end{equation}
where $dN_\gamma/dE$ is the differential gamma-ray spectrum per annihilation.
%where $M_{DM}$,$\langle{\sigma}v\rangle$ is mass and velocity averaged cross section of DM respectively.
The second term in Eq.(\ref{equ_dmflux}) is the so-called $J$-factor, which is governed by the DM density distribution $\rho(r)$ and reads
\begin{equation}
J_{\rm anni} = \int_\Omega\int_{s=0}^\infty{\rho^2(r(s))}dsd\Omega.
\end{equation}
 Clearly, the expected gamma-ray flux is proportional to $J$-factor, and the sources with larger $J$-factors have better potential to display DM signal.\\

In Table 1 we list all dSphs detected so far (including some recently discovered candidates) and their $J$-factors (if available).
The confirmation of a candidate to be a dSph and the determination of the amount of DM content (and then determining the $J$-factor) require spectroscopic measurements of velocities of a group of member stars. The dSph candidates listed in Table 1 are mainly discovered by the DES and Pan-STARRS1 and have not been spectroscopically observed yet. Some confirmed dSphs do not have reliable $J$-factors due to the lack of measurements of sufficient member stars or tidal stripings which makes the results uncertain.
Serval groups have been concentrated on evaluating $J-$factors systematically. Currently their results are not fully consistent with each other.
 To be less biased, in this work we adopt the $J$-factors from several different studies \cite{fermi14dsph,fermi2016dsph,gs15jfs,bonnivard15jfs,sanders16jfs}.

\begin{table}[ht]%[H] add [H] placement to break table across pages
\small
\caption{Parameters of the dSphs and candidates.}
\begin{ruledtabular}
\begin{tabular}{lrrcccccc}%{c||cc|cc|cc|cc|cc} %c means center column...l/r for left/rigth justify
%Event Type & SOURCE & CLEAN & ULTRACLEAN \\
%\multicolumn{1}{c}{} & \multicolumn{2}{c}{SOURCE} & \multicolumn{2}{c}{CLEAN} & \multicolumn{2}{c}{ULTRACLEAN} & \multicolumn{2}{c}{ULTRACLEANVETO}\\
Name & Longitude &  Latitude  & Distance & \multicolumn{5}{c}{$J$-factors}\\
%Name & Longitude &  Latitude  & Distance & $J$-factors1\footnote{Ackermann et al. 2014 \cite{fermi14dsph}}  & $J$-factors2\footnote{Geringer-Sameth et al. 2015 \cite{gs15jfs}} & $J$-factors3\footnote{Bonnivard et al. 2016 \cite{bonnivard15jfs}} & $J$-factors4\footnote{Sanders et al. 2016 \cite{sanders16jfs}} & $J$-factors5\footnote{Albert et al. 2016 \cite{fermi2016dsph}} \\
%\hline
  & [deg] &  [deg]  & [kpc] & \multicolumn{5}{c}{[GeV$^2$cm$^{-5}]$}\\
\hline
  &       &         &       & \cite{fermi14dsph}  & \cite{gs15jfs} & \cite{bonnivard15jfs} & \cite{sanders16jfs} & \cite{fermi2016dsph}\\
\hline

%#            name   & ll      &  bb     &d[kc]    & _sph(0.5))                        & log10(J0.5)                 &log10(J(0.5deg))     Jfactor
          Bootes\ I  & 358.10  &  69.60  &  66   & $18.8\pm0.22$ & $18.24_{-0.37}^{+0.40}$ & $18.5_{-0.4}^{+0.6}$ & $17.06_{-0.38}^{+0.65}$ & 18.5 \\
         Bootes\ II  & 353.70  &  68.90  &  42   & $     -     $ & $        -            $ & $       -          $ & $         -           $ & 18.9 \\
        Bootes\ III  & 35.40   &  75.40  &  47   & $     -     $ & $        -            $ & $       -          $ & $         -           $ & 18.8 \\
 Canes\ Venatici\ I  & 74.30   &  79.80  & 218   & $17.7\pm0.26$ & $17.43_{-0.28}^{+0.37}$ & $17.5_{-0.2}^{+0.4}$ & $17.68_{-0.11}^{+0.11}$ & 17.4 \\
Canes\ Venatici\ II  & 113.60  &  82.70  & 160   & $17.9\pm0.25$ & $17.65_{-0.43}^{+0.45}$ & $18.5_{-0.9}^{+1.2}$ & $18.06_{-0.40}^{+0.40}$ & 17.7 \\
       Canis\ Major  & 240.00  & -8.00   &   7   & $     -     $ & $        -            $ & $       -          $ & $         -           $ &  -   \\
             Carina  & 260.10  & -22.20  & 105   & $18.1\pm0.23$ & $17.87_{-0.09}^{+0.10}$ & $17.9_{-0.1}^{+0.2}$ & $18.40_{-0.34}^{+0.34}$ & 18.1 \\
    Coma\ Berenices  & 241.90  &  83.60  &  44   & $19.0\pm0.25$ & $19.02_{-0.41}^{+0.37}$ & $19.6_{-0.7}^{+0.8}$ & $19.08_{-0.32}^{+0.32}$ & 18.8 \\
              Draco  & 86.40   &  34.70  &  76   & $18.8\pm0.16$ & $18.84_{-0.13}^{+0.12}$ & $19.1_{-0.2}^{+0.4}$ & $19.27_{-0.24}^{+0.24}$ & 18.3 \\
          Draco\ II  & 98.30   &  42.90  &  24   & $     -     $ & $        -            $ & $       -          $ & $         -           $ & 19.3 \\
             Fornax  & 237.10  & -65.70  & 147   & $18.2\pm0.21$ & $17.83_{-0.06}^{+0.12}$ & $17.7_{-0.1}^{+0.1}$ & $18.56_{-0.16}^{+0.16}$ & 17.8 \\
           Hercules  & 28.70   &  36.90  & 132   & $18.1\pm0.25$ & $16.86_{-0.68}^{+0.74}$ & $17.5_{-0.7}^{+0.7}$ & $17.24_{-0.45}^{+0.45}$ & 17.9 \\
          Hydra\ II  & 295.61  &  30.46  & 134   & $     -     $ & $        -            $ & $       -          $ & $16.97_{-1.84}^{+0.87}$ & 17.8 \\
      Horologium\ I  & 271.40  & -54.70  &  87   & $     -     $ & $        -            $ & $       -          $ & $19.05_{-0.39}^{+0.95}$ & 18.2 \\
             Leo\ I  & 226.00  &  49.10  & 254   & $17.7\pm0.18$ & $17.84_{-0.16}^{+0.20}$ & $17.8_{-0.2}^{+0.5}$ & $18.21_{-0.28}^{+0.28}$ & 17.3 \\
            Leo\ II  & 220.20  &  67.20  & 233   & $17.6\pm0.18$ & $17.97_{-0.18}^{+0.20}$ & $18.0_{-0.2}^{+0.6}$ & $17.85_{-0.25}^{+0.25}$ & 17.4 \\
            Leo\ IV  & 265.40  &  56.50  & 154   & $17.9\pm0.28$ & $16.32_{-1.69}^{+1.06}$ & $16.2_{-1.6}^{+1.3}$ & $17.05_{-0.91}^{+0.90}$ & 17.7 \\
             Leo\ V  & 261.90  &  58.50  & 178   & $     -     $ & $16.37_{-0.87}^{+0.94}$ & $16.1_{-1.0}^{+1.2}$ & $17.35_{-0.71}^{+1.06}$ & 17.6 \\
             Leo\ T  & 214.85  &  43.66  & 407   & $     -     $ & $17.11_{-0.39}^{+0.44}$ & $17.6_{-0.6}^{+1.0}$ & $17.73_{-0.37}^{+0.38}$ &  -   \\
         Pisces\ II  & 79.20   & -47.10  & 182   & $     -     $ & $        -            $ & $       -          $ & $18.30_{-0.80}^{+1.15}$ & 17.6 \\
      Reticulum\ II  & 266.30  & -49.70  &  32   & $     -     $ & $        -            $ & $       -          $ & $19.12_{-0.33}^{+0.85}$ & 19.1 \\
        Sagittarius  & 5.60    & -14.20  &  26   & $     -     $ & $        -            $ & $       -          $ & $         -           $ &  -   \\
           Sculptor  & 287.50  & -83.20  &  86   & $18.6\pm0.18$ & $18.54_{-0.05}^{+0.06}$ & $18.5_{-0.1}^{+0.1}$ & $19.07_{-0.29}^{+0.29}$ & 18.2 \\
           Segue\ 1  & 220.50  &  50.40  &  23   & $19.5\pm0.29$ & $19.36_{-0.35}^{+0.32}$ & $17.0_{-2.2}^{+2.1}$ & $19.81_{-0.39}^{+0.39}$ & 19.4 \\
           Segue\ 2  & 149.40  & -38.10  &  35   & $     -     $ & $16.21_{-0.98}^{+1.06}$ & $18.9_{-1.1}^{+1.1}$ & $17.52_{-1.74}^{+0.86}$ &  -   \\
            Sextans  & 243.50  &  42.30  &  86   & $18.4\pm0.27$ & $17.52_{-0.18}^{+0.28}$ & $17.6_{-0.2}^{+0.2}$ & $18.28_{-0.29}^{+0.29}$ & 18.2 \\
     Triangulum\ II  & 141.40  & -23.40  &  30   & $     -     $ & $        -            $ & $       -          $ & $         -           $ & 19.1 \\
         Tucana\ II  & 328.00  & -52.40  &  58   & $     -     $ & $        -            $ & $       -          $ & $19.45_{-0.58}^{+0.87}$ & 18.6 \\
     Ursa\ Major\ I  & 159.40  &  54.40  &  97   & $18.3\pm0.24$ & $17.87_{-0.33}^{+0.56}$ & $18.7_{-0.4}^{+0.6}$ & $18.89_{-0.25}^{+0.25}$ & 18.1 \\
    Ursa\ Major\ II  & 152.50  &  37.40  &  32   & $19.3\pm0.28$ & $19.42_{-0.42}^{+0.44}$ & $19.9_{-0.5}^{+0.7}$ & $19.78_{-0.39}^{+0.39}$ & 19.1 \\
        Ursa\ Minor  & 105.00  &  44.80  &  76   & $18.8\pm0.19$ & $18.93_{-0.19}^{+0.27}$ & $19.0_{-0.1}^{+0.1}$ & $19.56_{-0.24}^{+0.24}$ & 18.3 \\
         Willman\ 1  & 158.60  &  56.80  &  38   & $19.1\pm0.31$ & $        -            $ & $19.5_{-0.6}^{+1.2}$ & $19.69_{-0.62}^{+0.92}$ & 18.9 \\
\hline
          Cetus\ II  & 156.48  & -78.53  &  30   & $     -     $ & $        -            $ & $       -          $ & $         -           $ & 19.1 \\
         Columba\ I  & 231.62  & -28.88  & 182   & $     -     $ & $        -            $ & $       -          $ & $         -           $ & 17.6 \\
      Eridanus\ III  & 275.00  & -59.60  &  96   & $     -     $ & $        -            $ & $       -          $ & $         -           $ & 18.1 \\
       Eridanus\ II  & 249.80  & -51.60  & 331   & $     -     $ & $        -            $ & $       -          $ & $         -           $ & 17.1 \\
            Grus\ I  & 338.68  & -58.24  & 120   & $     -     $ & $        -            $ & $       -          $ & $18.35_{-1.92}^{+0.92}$ & 17.9 \\
           Grus\ II  & 351.15  & -51.94  &  53   & $     -     $ & $        -            $ & $       -          $ & $         -           $ & 18.7 \\
     Horologium\ II  & 262.47  & -54.14  &  78   & $     -     $ & $        -            $ & $       -          $ & $         -           $ & 18.3 \\
          Indus\ II  & 353.99  & -37.40  & 214   & $     -     $ & $        -            $ & $       -          $ & $         -           $ & 17.4 \\
       Pegasus\ III  & 69.85   & -41.81  & 205   & $     -     $ & $        -            $ & $       -          $ & $         -           $ & 17.5 \\
        Pictoris\ I  & 257.30  & -40.60  & 126   & $     -     $ & $        -            $ & $       -          $ & $         -           $ & 17.9 \\
        Phoenix\ II  & 323.70  & -59.70  &  96   & $     -     $ & $        -            $ & $       -          $ & $         -           $ & 18.1 \\
     Reticulum\ III  & 273.88  & -45.65  &  92   & $     -     $ & $        -            $ & $       -          $ & $         -           $ & 18.2 \\
    Sagittarius\ II  & 18.90   & -22.90  &  67   & $     -     $ & $        -            $ & $       -          $ & $         -           $ & 18.4 \\
        Tucana\ III  & 315.38  & -56.19  &  25   & $     -     $ & $        -            $ & $       -          $ & $         -           $ & 19.3 \\
         Tucana\ IV  & 313.29  & -55.29  &  48   & $     -     $ & $        -            $ & $       -          $ & $         -           $ & 18.7 \\
          Tucana\ V  & 316.31  & -51.89  &  55   & $     -     $ & $        -            $ & $       -          $ & $         -           $ & 18.6 \\
\end{tabular}
\end{ruledtabular}
\footnotetext{{\it Note}: The summary of dSphs and candidates (divided by a horizontal line) published in the literature.
%The sources above the horizontal line are the ones have been confirmed as dSphs by spectroscopic measurements, while below the horizontal line are dSph candidates.
 The listed $J$-factors are from \cite{fermi14dsph,fermi2016dsph,gs15jfs,bonnivard15jfs,sanders16jfs}, respectively. The divergency between them suggests that the $J$-factors adopted in different literature likely suffer from non-ignorable systematic uncertainty.
The $J$-factors in \cite{sanders16jfs} were calculated using a simple analytical method rather than Markov Chain Monte Carlo based Jeans analysis. The $J$-factors in the last column were estimated by assuming there is a scaling relationship between $J$-factor and distance \cite{des_fermi15dsph,fermi2016dsph}. Note that Grus I has been spectroscopically observed already and a dwarf galaxy nature is favored, nevertheless further velocity dispersion measurements are acquired to draw the final conclusion \cite{walker16grus1}.}
\label{tb1}
\end{table}

\section{Searching for line like signals towards dSphs}
\label{sec_dsph}
\subsection{Data Selection}
The Fermi-LAT \cite{atwood09lat} is a pair conversion instrument sensitive for gamma-ray detection in the energy range $\sim$30~MeV to {\textgreater}500~GeV. In this work, the publicly released Pass 8 data (P8R2 Version 6) from the Fermi-LAT are analyzed \footnote{\url{http://fermi.gsfc.nasa.gov/ssc/data/access/}}. The Pass 8 data release provides a number of improvements compared to previous versions, including a wider energy range, better energy measurements, and a significantly increased effective area. In addition, the Pass 8 data are subdivided into quartiles according to events' energy/direction reconstruction qualities, allowing to improve the energy/direction resolution by using the high quality data only \cite{pass8econf}.
We take into account 91 months (from 2008-10-27 to 2016-06-08, i.e. MET 246823875 - MET 487121910)
of data, with energies between 1 and 500 GeV.
The zenith-angle cut $\theta < 90^\circ$ is applied in order to filter out the Earth's limb emission which is a strong source of background contamination. We adopt the recommended quality-filter cuts (DATA\_QUAL{\textgreater}0 \&\& LAT\_CONFIG==1) to extract the good time intervals.
Throughout the work, we make use of the ULTRACLEAN data set in order to reduce the contamination from charged cosmic rays. Since the energy resolution of EDISP0 data is much worse than that of the rest and it just accounts for ${\sim}$1/4 of the whole data sets\footnote{\url{http://www.slac.stanford.edu/exp/glast/groups/canda/lat_Performance.htm}}, following L16 we exclude the EDISP0 data in our analysis to achieve better energy resolution without significant loss of the statistics. For each dSph we utilize {\it gtselect} to select data within 1 degree radius. We generate HEALPIX format exposure map using the {\it gtexpcube2} tool. The selection of events as well as the calculation of exposure maps are performed with the latest v10r0p5 version of Fermi science tools.

\subsection{Data Analysis}
We use a stacked, unbinned likelihood analysis together with the {\it sliding energy windows} technique \cite{bringmann130gev, weniger130gev,pullen07egret} (see also L16) to search for the line-like signals from the dSphs (and candidates).
Events within all of our ROIs are gathered together and divided by exposures averaged over all ROIs to yield a stacked spectrum. Then this stacked spectrum is fitted using an unbinned maximum-likelihood method for a series of  $E_{\gamma}$ from 5~GeV to 300~GeV with increment in steps of 0.5 ${\sigma}_E(E_{\gamma})$, where $E_{\gamma}$ is the energy of putative line signal which is fixed in the fitting procedure and ${\sigma}_E(E_{\gamma})$ is the energy resolution ($68{\%}$ containment) of the LAT at $E_{\gamma}$. For each $E_{\gamma}$, the fitting is performed in a small energy window of $(E_{\gamma}-0.5E_{\gamma},~E_{\gamma}+0.5E_{\gamma})$. This small window size warrants that the background spectrum can be well approximated as a power law.
We use two models to fit the spectrum in each energy window: 1) single power law, 2) power law background plus a line signal. Considering the energy dispersion, the line component is expressed as exposure weighted Fermi-LAT energy dispersion function\footnote{\url{http://fermi.gsfc.nasa.gov/ssc/data/analysis/documentation/Cicerone/Cicerone_LAT_IRFs/IRF_E_dispersion.html}}.
The test statistics (TS) of a line component can be derived by comparing the likelihood values between these two models,
\begin{equation}
{\rm TS}\triangleq -2\ln\frac{\mathcal{L}_{\rm model1}}{\mathcal{L}_{\rm model2}}.
\end{equation}
The local significance is just the square root of TS, since the TS value would follow $\chi^2$ distribution with only one degree of freedom according to the asymptotic theorem of Chernoff \cite{chernoff}. Taking into account a trial factor would further decrease the significance.
The analysis procedures are the same as L16. For simplicity we do not re-introduce them here and refer the readers to Sec. II B, Sec. II C, and Sec. III A of L16 for the stacking method, unbinned likelihood method and sliding window technique, respectively.

\subsection{Results}
Our search results are presented in the left panel of Fig.1, which exhibits how the TS value of a putative line-like signal varies as a function of the line energy. Clearly, no significant line signal is found over all the energy range we consider (i.e., there is no signal with ${\rm TS}>9$, corresponding to a local significance of $3{\sigma}$).
Note that this plot is for 48 ROIs including both the confirmed dSphs and the candidates. As an independent check, in the right panel of Fig.1 we present the result for 32 confirmed dSphs only. No significant signal is found either.

\begin{figure}[!h]
\includegraphics[width=0.45\columnwidth]{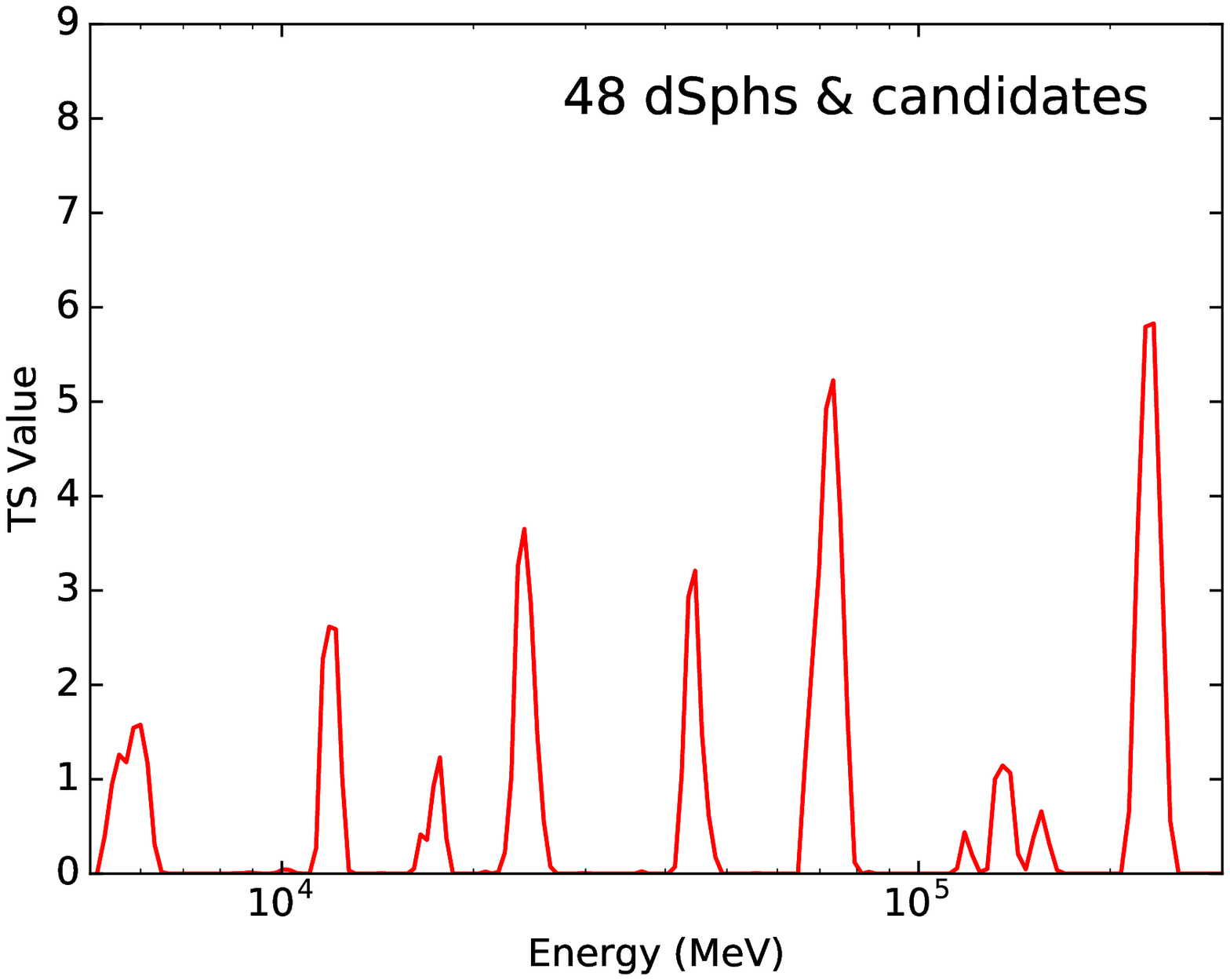}
\includegraphics[width=0.45\columnwidth]{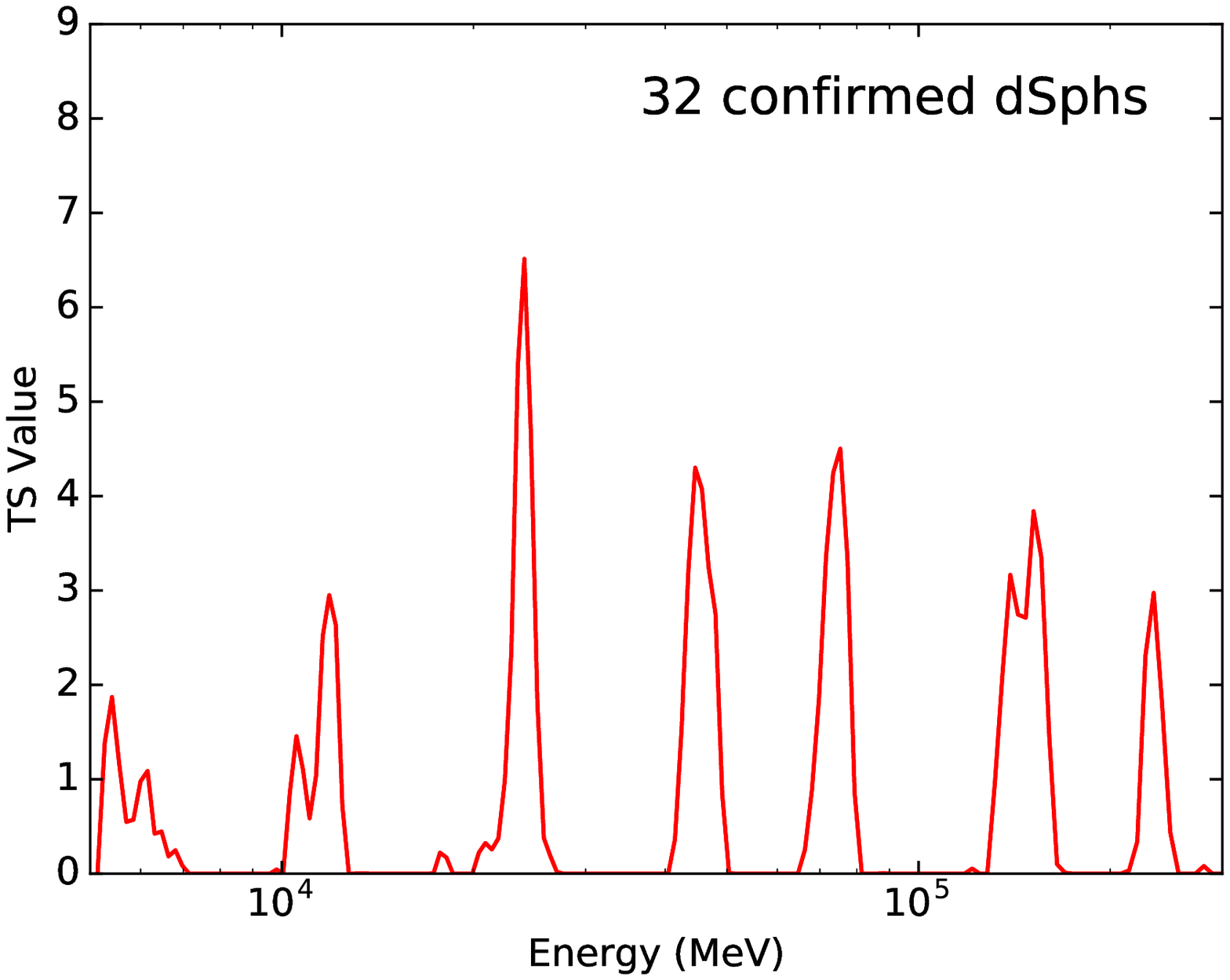}
\caption{The TS value as a function of the line energy in the sliding window analysis. The left panel is for the 48 confirmed dSphs and candidates, and the right panel is for the 32 confirmed dSphs only. No significant signal is found over all the energies we have examined.}
\label{slide}
\end{figure}

Due to the absence of any significant line-like signals from the dSphs, we set constraints on the cross section of DM annihilating into two photons $\langle \sigma v \rangle_{\chi\chi\rightarrow \gamma\gamma}$. For DM annihilating into a pair of $\gamma$-rays (i.e., $\chi\chi\to\gamma\gamma$), the expected flux is given by
\begin{equation}
S_{\rm line}(E) = \frac{1}{4\pi} \frac{\left<\sigma v\right>_{\chi\chi\to\gamma\gamma}}{2m_\chi^2} \ 2\delta(E-E_{\rm line})\sum_{\rm i=1}^{n} J_{\rm i},
\label{eg:flux}
\end{equation}
where %$m_\chi$ is the rest mass of the DM particle, $\left<\sigma v\right>_{\chi\chi\to\gamma\gamma}$ is the velocity-averaged annihilation cross-section for $\chi\chi\to\gamma\gamma$,
$E_{\rm line}=m_\chi$ is the energy of the monoenergetic photons. We sum the $J$-factors over all the objects we have analyzed.

\begin{figure}[!h]
\includegraphics[width=0.5\columnwidth]{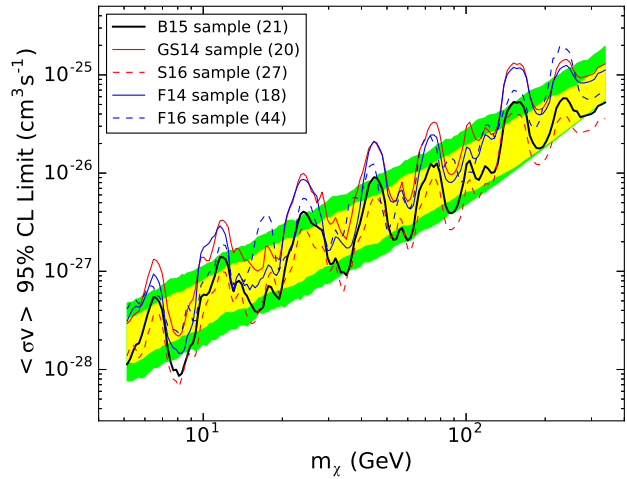}
\caption{The 95\% confidence level upper limits on the cross section of DM annihilating into double $\gamma$-rays obtained in our analysis of the dSphs (black solid line). Yellow and green bands represent 68\% and 95\% containment of limits obtained by $10^{3}$ blank-sky simulations, respectively.
%We notice that at the high energy end, the lower edges of 68\% and 95\% containment bands become superposing each other. This is because the distribution of $10^3$ simulated upper limits is no longer a Gaussian distribution due to the limited event number. 
Also plotted are results based on other samples with different $J$-factors. Note that they can not be compared with neither the ``fiducial" result (black solid line) nor blank sky band directly since the numbers of dSphs contained in different samples are not the same. The bracketed number in the line label represents the number of dSphs in the given sample.}
\label{limits}
\end{figure}

Following L16, for a given $m_\chi$ we fit the data in the corresponding energy window with a series of $\left<\sigma v\right>_{\chi\chi\to\gamma\gamma}$ in Eq.(\ref{eg:flux}) and find the cross section at which the
log-likelihood is smaller by 1.35 compared to the maximum one, which corresponds to the 95\% confidence level upper limit of the cross section.
Unlike the search for excess in Fig.\ref{slide}, for constraining $\left<\sigma v\right>_{\chi\chi\to\gamma\gamma}$ the information of $J$-factors is necessary to convert gamma-ray flux limits into DM annihilation cross section. Therefore, our current sample is not the same as that adopted in Fig.\ref{slide}. We use the 21 dSphs with $J$-factors reported by \cite{bonnivard15jfs} to derive the ``fiducial" constraints.
Fig.~\ref{limits} shows the resulting constraints on $\langle \sigma v \rangle_{\chi\chi\rightarrow \gamma\gamma}$ in the energy range 5~GeV$-$300~GeV. Also shown are the expected 68\% and 95\% containment regions derived from $10^{3}$ blank-sky Monte Carlo simulations (the yellow and green bands).
In each simulation, a set of 21 high-latitude blank-sky ROIs non-overlapping with 21 dSphs are selected and analyzed with the same procedure as stated above to derive upper limits. 
We notice that at the high energy end, the lower edges of 68\% and 95\% containment bands become superposing each other. This is because the distribution of $10^3$ simulated upper limits is no longer a Gaussian distribution due to the limited event number.

In Fig.~\ref{limits} we also present the constraints based on samples with $J$-factors derived by other groups (see Sec.\ref{samples} for details of these samples), with blue and red solid/dash lines. One should note that these results can not be directly compared with our ``fiducial" results (black line) since the numbers of dSphs in these samples are not the same. We just plot them here as a reference to show how significant the upper limits will change in these scenarios. For the same reason, the 68\% and 95\% containment regions (the yellow and green bands) are only valid for the ``fiducial" constraints.

\begin{figure}[!h]
\includegraphics[width=0.5\columnwidth]{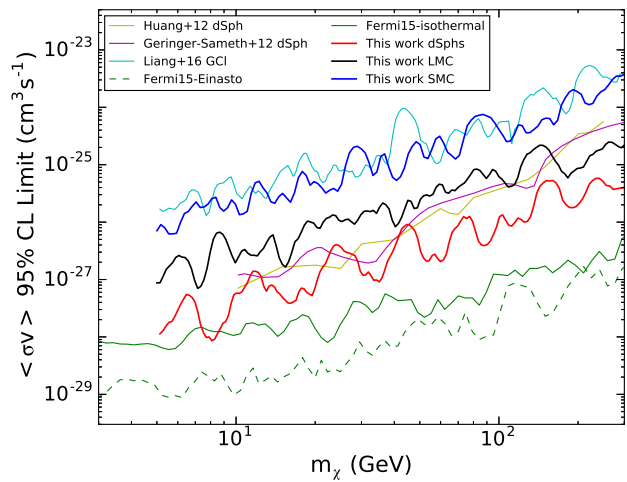}
\caption{A compilation of constraints on cross section of the DM annihilating to double $\gamma$-rays. Our ``fiducial" result obtained by analyzing data from 21 dSphs is shown in red. The results from the analyses of the LMC and SMC data are shown in black and blue. For comparison, we also plot several constraints from other studies. The cyan line is for the constraints set by galaxy clusters \cite{liang16line}. The green solid/dashed lines are the constraints set by the Galactic gamma-ray data in the case of Isothermal and Einasto DM distributions, respectively \cite{fermi15line5}. The yellow and magenta lines are constraints for dSphs reported by Huang et al. \cite{huang12130gev} and by Geringer-Sameth et al. \cite{gs12dsphline}, respectively.
}
\label{compilation}
\end{figure}

In Fig.\ref{compilation} we compare our limits with some previous independent constraints. The thick red line is our ``fiducial" result. The cyan line represents the constraints set by galaxy clusters (adopted from L16 and without the boost factor correction). The two green lines are for the constraints from Galactic gamma-ray data, where the dashed line is for Einasto DM distribution while the solid line is for the Isothermal DM distribution \cite{fermi15line5}. The yellow line and magenta line are the constraints set by Pass 7 dSph data \cite{huang12130gev,gs12dsphline}. Interestingly, though for continual signal the dSphs yield the most stringent constraints, for line-like signal the constraints by dSphs are weaker than that set by Galactic gamma-ray data notably (the dSph constraints are one or more orders of magnitude weaker than the Galactic data constraints).
This indicates that, for DM line searches a larger $J$-factor(s) is more important than cleaner background due to the distinctive spectrum feature of the signal. The Galactic center is a better target unless much more dSphs with larger $J$-factor are found in the future. For the same reason, the galaxy clusters give the weakest constrained unless the boost factors are very high.

\section{Searching for line-like signals towards LMC \& SMC}
\label{lmcsmc}
The Magellanic Clouds, the largest satellite galaxies of the Milky Way, are also natural targets for the DM indirect detection searches. In particular, the LMC is expected to be the second brightest source of gamma rays from the DM annihilation in the sky£¬ due to its ${\sim}50$ kpc distance and ${\sim}10^{10} M_\odot$ predicted DM mass.  Prior to this work, search for gamma-ray emission from the DM annihilation in the LMC has been performed for the $s\bar{s}$, $b\bar{b}$, $t\bar{t}$, $gg$, $W^+W^-$, $e^+e^-$, $\mu^+\mu^-$ and $\tau^+\tau^-$ channels \cite{buckley15lmc}. A similar analysis of the SMC has also been performed in \cite{caputo16smc}. Unlike the previous analysis, here we perform a line-search for the DM annihilating directly into photons in the LMC/SMC.

We take the relatively conservative value of $J=9.4\times10^{19}$~GeV$^2$/cm$^5$ for the LMC \cite{buckley15lmc} and $J=1.13\times10^{19}$~GeV$^2$/cm$^5$ for the SMC \cite{caputo16smc}. These $J$-factors were computed by integrating to an angular distance of 15$^\circ$ from the center (assuming an NFW spatial profile \cite{nfw96nfw}). For the LMC, the characteristic density is assumed to be $\rho_0=2.6\times10^6~M_\odot$/kpc$^3$ and the scale radius is assumed to be $r_s=12.6$~kpc. For the SMC, we have $\rho_0=4.1\times10^6~M_\odot$/kpc$^3$ and $r_s=5.1$~kpc.

The ROIs are defined as a circle with 15$^\circ$ radius centered on ${\rm (RA,~DEC)}=(80^{\circ}.89,~-69^{\circ}.76)$ for the LMC, and $(13^{\circ}.16,~-72^{\circ}.80)$ for the SMC, respectively.
Other data selection criterion and the line-search procedure are the same as that in Sec.\ref{sec_dsph}.

No significant line signal in the energy range of $5-300$ GeV is found for these two sources either. The largest TS appears at $\sim$8.5~GeV for the LMC (with a TS$\sim$8) and there is no other spectral structure with TS$>$4. For the SMC the largest TS is $\sim5$ at $\sim$28~GeV. Since no signal is found, we place 95\% upper limits on $\langle \sigma v \rangle_{\chi\chi\rightarrow \gamma\gamma}$. The limits are plotted in Fig.\ref{compilation} as black for the LMC and blue for the SMC. Despite the $J$-factor of the LMC is comparable with that of some dSphs ($9.4\times10^{19}$~GeV$^2$/cm$^5$ for the LMC vs. $20.2\times10^{19}$~GeV$^2$/cm$^5$ for our ``fiducial" dSph sample), the limits obtained by the former is several times weaker (note that the time interval and the type of the data are the same as that adopted in the dSphs analysis). This is mainly because the LMC has higher background emission originating from interactions between cosmic rays and the interstellar medium and from point sources such as pulsars within the LMC and especially within the 30 Doradus star-forming region \cite{fermi16lmc}.

\section{Summary and Discussion}
A robust detection of a monochromatic gamma-ray line would serve as a smoking-gun to prove the existence of particle DM. That is why great effects have been made to search for such signals in various targets since Fermi's successful launch in 2008. In this work, we have analyzed 91 months' publicly-available Pass 8 Fermi-LAT data in the directions of a sample of dwarf spheroidal galaxies (including candidates) and LMC/SMC. Our search results show no significant signals in the energy range  5~$-$300~GeV. The gamma-ray data are well consistent with the background only blank sky simulations. Thus we set limits on the DM annihilation cross section to produce monochromatic gamma rays. Comparing with the constraints set by the 4 years Pass 7 Fermi-LAT data in the directions of seven dSphs, our current limits are much tighter in a wide energy range. However our limits are still weaker than the constraints set by the Galactic gamma-ray data even for an ``isothermal" DM distribution profile \cite{fermi15line5}. This may indicate that for DM line searches, a larger $J$-factor is more important than cleaner background due to the distinctive spectrum characteristic of the signal. However, the situation may be changed with the fast developing dSph surveys. The DES collaboration has already found 16 dSphs (including candidates) in their first two years' searches and much more are expected to be identified in the upcoming years \cite{des15y1,des15y2}. In the near future, the LSST \cite{ivezic08lsst} is expected to discover hundreds of new dSphs. With remarkably growing sample of dSphs, the sensitivity of searching for the DM signal (including both the line search as well as the continuum emission search) will get improved significantly.

Finally, we would like to point out that China's operating space mission, the Dark Matter Particle Explorer \cite{Chang2014}, and a proposing future mission, the High Energy cosmic-Radiation Detection Facality \cite{zhang14herd}, which are dedicated to measuring high-energy cosmic ray electrons and gamma rays with the unprecedentedly high energy resolution in a wide energy range, are expected to contribute significantly to the gamma-ray line search \cite{huang2015herd}.
%Finally, we would like to point out that the Dark Matter Particle Explorer \cite{Chang2014}, a Chinese space mission dedicated to measure high-energy cosmic ray electrons and gamma rays with the unprecedented energy resolution in a wide energy range, is expected to contribute significantly to the gamma-ray line search.

\begin{acknowledgments}
This work was supported in part by the National Basic Research Program of China (No. 2013CB837000), National Natural Science Foundation of China under grants No. 11525313 (i.e., the Funds for Distinguished Young Scholars) and No. 11103084, and by the Strategic Priority Research Program (No. XDA04075500).
\end{acknowledgments}

%\newpage

\end{document}